\documentclass[runningheads]{llncs}
\usepackage[T1]{fontenc}
\usepackage{float}
\usepackage{amsmath}
\usepackage{amsfonts}
\usepackage{graphicx}
\usepackage{booktabs}
\usepackage{tabularx} 
\usepackage{makecell}
\usepackage{layouts}
\begin{document}
\title{Modulated INR with Prior Embeddings for Ultrasound Imaging Reconstruction}
\titlerunning{Modular INR for Ultrafast Ultrasound Imaging}

\author{R\'emi Delaunay \and Christoph Hennersperger \and Stefan Wörz}

\authorrunning{R. Delaunay et al.}

\institute{LUMA Vision GmbH, Munich, Germany \\
\email{remi.delaunay@lumavision.com}}
    
\maketitle              

\begin{abstract}
Ultrafast ultrasound imaging enables visualization of rapid physiological dynamics by acquiring data at exceptionally high frame rates. However, this speed often comes at the cost of spatial resolution and image quality due to unfocused wave transmissions and associated artifacts. In this work, we propose a novel modulated Implicit Neural Representation (INR) framework that leverages a coordinate-based neural network conditioned on latent embeddings extracted from time-delayed I/Q channel data for high-quality ultrasound image reconstruction. Our method integrates complex Gabor wavelet activation and a conditioner network to capture the oscillatory and phase-sensitive nature of I/Q ultrasound signals. We evaluate the framework on an \textit{in vivo} intracardiac echocardiography (ICE) dataset and demonstrate that it outperforms the compared state-of-the-art methods. We believe these findings not only highlight the advantages of INR-based modeling for ultrasound image reconstruction, but also point to broader opportunities for applying INR frameworks across other medical imaging modalities.
\end{abstract}

\section{Introduction}
Ultrafast ultrasound imaging, which achieves frame rates exceeding 1,000 frames per second, holds significant clinical potential in the assessment of cardiac function \cite{villemain2020ultrafast}. In contrast to conventional pulse echo acquisitions that are performed using focused transmit beams for each image scan line, ultrafast imaging relies on the insonification of the entire field of view by transmitting a single unfocused wavefront, such as a plane wave (PW) or a diverging wave (DW) \cite{tanter2014ultrafast}.  However, the absence of transmit focusing also results in lower image resolution and more pronounced ultrasound artifacts, such as grating lobes and side lobes. Techniques like coherent plane-wave compounding (CPWC) can be used to reduce these artifacts by combining multiple steered unfocused acquisitions, though at the cost of a reduced frame rate~\cite{montaldo2009coherent}.

Neural networks have shown great potential in addressing the trade-off between image quality and frame rate in ultrasound imaging. High-quality reconstruction techniques typically train neural networks to learn a non-linear mapping between low-quality images reconstructed from a small number of unfocused plane wave transmissions and high-quality reference images. Convolutional neural networks (CNN) have been employed to enhance both post-beamformed radio frequency (RF) signals \cite{perdios2021cnn} and in-phase/quadrature (I/Q) data \cite{lu2021complex,vinals2024enhancement}. In parallel, generative AI methods have been applied to PW imaging to produce high-quality reconstructions \cite{tang2021plane,asgariandehkordi2024denoising}. CNNs have also been incorporated into the beamforming process itself by applying convolutional layers to time-delayed RF channel data and directly reconstruct B-mode images \cite{xiao2025beamforming}. Additionally, feedforward neural networks have been used to process time-delayed RF data and predict content-adaptive apodization weights, exploiting the algorithmic structure of minimum variance (MV) beamforming~\cite{luijten2020adaptive}. \\
\indent Implicit neural representations (INRs) have emerged as a promising alternative to traditional grid-based data representations, offering compact and continuous representations of data, enabling tasks like image reconstruction and compression with high accuracy \cite{Molaei_2023_ICCV}. However, traditional INRs often struggle to capture fine, high-frequency details, a limitation largely attributed to the smoothness bias introduced by conventional activation functions (e.g., ReLU) \cite{ramasinghe2022gaussian}. To address this, various positional encoding techniques \cite{tancik2020fourierrelupe,mildenhall2021nerf} and alternative activation functions, such as sinusoidal activations \cite{sitzmann2020siren,liu2024finer}, Gabor wavelet activations \cite{saragadam2023wire}, and Gaussian activation \cite{ramasinghe2022gaussian}, have been proposed to improve the modeling of intricate structures. Another limitation of INRs lies in their reduced ability to generalize across varying inputs, which has led to an increased investigation of conditional architectures. These models adapt the INR based on external inputs, either through hypernetworks, auxiliary networks that generate INR weights dynamically \cite{chen2022transformers,sen2023hyp}, or latent embeddings, which encode instance-specific information to condition the INR~\cite{mehta2021modulated,kazerouni2024incode}.\\
\indent In this work, we propose a novel modulated INR framework for high-quality reconstruction in ultrafast ultrasound imaging. In contrast with prior ultrasound reconstruction approaches which operate on grid-based representations, our method provides a continuous, compact representation that can model high-quality images from a limited number of unfocused transmissions. Furthermore, while conventional INRs developed for vision tasks struggle to adapt to varying inputs due to their static nature, our conditional design enables the INR to generalize across different ultrasound acquisitions by modulating the network with instance-specific latent embeddings. Our MLP architecture also adopts complex-valued weights and nonlinear activations to better capture the oscillatory and phase-sensitive characteristics of I/Q ultrasound signals. \\ 
\indent We evaluate our approach on a dedicated \textit{in vivo} intracardiac echocardiography (ICE) porcine dataset, showing higher quantitative scores over state-of-the-art methods. An ablation study further highlights the contribution of each component. By leveraging conditional modeling and complex-valued representations, our method offers a compact and adaptive INR framework for high-quality ultrasound image reconstruction.

\section{Methods}
\subsection{Ultrasound image reconstruction}

One of the most commonly employed and well-established methods for reconstructing medical ultrasound images from raw channel data is delay-and-sum (DAS) beamforming \cite{perrot2021so}. It operates by temporally aligning and summing signals received at different transducer channels to estimate the echo intensity at each spatial location. For a given focal point \( \mathbf{r} \) in the imaging region, the round-trip propagation delay from the transmit origin \( \mathbf{r}_{\text{tx}} \) to the point \( \mathbf{r} \), and then to the \( i \)-th receive channel \( \mathbf{r}_i \), is computed under the assumption of a constant speed of sound \( c \). The total delay for the \( i \)-th channel is given by

\begin{equation}
\label{eq1}
t_i(\mathbf{r}) = \frac{\|\mathbf{r} - \mathbf{r}_{\text{tx}}\| + \|\mathbf{r} - \mathbf{r}_i\|}{c}
\end{equation}

This delay aligns the received signals and is commonly referred to as time-of-flight correction. To reduce sidelobe artifacts and improve image quality, each channel signal \(s_i \) can be weighted by an apodization coefficient \( w_i \), typically defined by a window function (e.g., rectangular, Hann, Tukey). The beamformed signal at location \( \mathbf{r} \) is then computed as

\begin{equation}
\label{eq2}
y(\mathbf{r}) = \sum_{i=1}^{N} w_i \cdot s_i(t_i(\mathbf{r}))
\end{equation}

This summation reinforces signals originating from \( \mathbf{r} \) through constructive interference while attenuating contributions from off-axis scatterers.

\subsection{Proposed framework}

INRs aim to approximate a continuous mapping \( S(x): \mathbb{R}^{D_i} \rightarrow \mathbb{C}^{D_o} \) by optimizing the parameters \( \theta \) of a MLP, \( f(x; \theta) \). In our setup, the input dimension \( D_i = 3\) corresponds to 2D spatial coordinates and the transmit steering angle. The network can either directly output the mapped complex I/Q value (\( D_o = 1 \)), or predict a set of \( N \) complex apodization weights (\( D_o = N \)), which is then applied to the time-delayed I/Q input (see Eq.~\ref{eq2}). The network is trained by minimizing

\begin{equation}
\arg\min_{\boldsymbol{\theta}} \mathop{\mathbb{E}}\limits_{\mathbf{x} \in \mathcal{X}} \left\| \mathcal{L}(f(\mathbf{x}; \boldsymbol{\theta})) - \mathcal{L}(S(\mathbf{x})) \right\|_2^2
\end{equation}

\noindent where \(\mathcal{L}(\cdot)\) denotes the B-mode conversion operation applied to both the predicted and ground truth signals, and \(\mathcal{X}\) represents the set of all spatial and steering angle coordinates sampled from the ground truth reconstructions used to supervise the INR. This loss formulation ensures that the network optimizes the perceptual quality of the ultrasound images by comparing their B-mode representations rather than the raw complex outputs.

\begin{figure}[H]
\centering
\includegraphics[width=\textwidth]{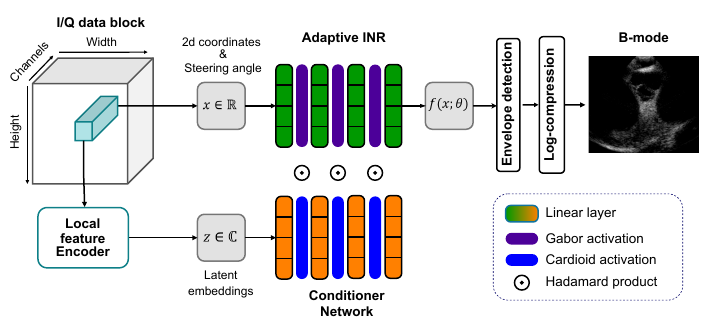}
\caption{Method overview.} \label{fig1}
\end{figure}

Our proposed framework consists of three key components: an adaptive INR, a modulation network, and a local prior encoder, which we describe below. The overall pipeline is shown in Fig.~\ref{fig1}.

\subsubsection{Adaptive INR network}

We define the Adaptive INR network as an MLP with L hidden layers, each containing P hidden features that map the input coordinates and steering angle to its output domain, i.e., I/Q signal values or apodization weights. Each hidden layer within the network comprises a complex Gabor wavelet $\psi$ as a nonlinear activation function \cite{saragadam2023wire}. The activation is defined as
\begin{equation}
\label{eqgabor}
\sigma(x) = \psi(x; \omega_0, s_0) = e^{j \omega_0 x} \cdot e^{- |s_0 x|^2},
\end{equation}

\noindent where \( \omega_0 \in \mathbb{R} \) controls the frequency of the modulation and \( s_0 \in \mathbb{R} \) controls the spatial spread (or width). This activation function offers several advantages in the context of I/Q data processing. First, it produces complex-valued outputs natively, preserving the inherent structure of the signal without needing real-imaginary separation. Second, the wavelet form enables localized frequency sensitivity, capturing spatially varying oscillatory features that are critical in applications such as high-quality image reconstruction \cite{sitzmann2020siren}. Finally, the non-saturating nature of the complex exponential ensures consistent gradient flow, and thus supports stable training.

\subsubsection{Local Prior Encoder Network}

The local prior encoder network is designed to be lightweight and efficient, serving the purpose of extracting a compact latent embedding \( \mathbf{z} \) from the time-delayed channel data. It is composed of three convolutional layers, each followed by a Cardioid activation function, and a final linear layer that maps the intermediate features to a low-dimensional latent space. The choice of the Cardioid activation is motivated by its suitability for complex-valued data, such as I/Q signals \cite{vinals2024enhancement}. Unlike standard activation functions like rectified linear unit (ReLU), which are designed for real-valued inputs, the Cardioid activation can operate on complex numbers while preserving important phase information \cite{barrachina2023theory,virtue2017better}. 

\subsubsection{Conditioner network}

The conditioner network consists of an MLP with \(L\) hidden layers, each followed by a Cardioid activation function, operating on the latent code \( \mathbf{z} \). A skip connection from \( \mathbf{z} \) to each hidden layer introduces an explicit dependency between the modulation parameters and the latent representation, similar to Mehta \textit{et al.}~\cite{mehta2021modulated}. The output of each hidden layer in the conditioner network is used to modulate the corresponding Gabor wavelet activation in the adaptive INR network. The activation at the \( i \)-th layer of the adaptive INR represented in Eq. \ref{eqgabor} then becomes
\begin{equation}
\mathbf{h}_i = \boldsymbol{\alpha}_i \odot \psi\left(W_i \cdot \mathbf{h}_{i-1} + \mathbf{b}_i; \omega_0, s_0\right),
\end{equation}
where \( \boldsymbol{\alpha}_i \) is the output of the \( i \)-th layer of the conditioner network. This formulation enables the conditioner network to directly modulate the amplitude of the signal representation at each layer.

\section{Experiments}

\subsection{\textit{In vivo} ICE Dataset}

The \textit{in vivo} dataset used in this study was acquired from porcine subjects using a custom-made intracardiac echocardiography (ICE) catheter equipped with 64 elements operating at a central frequency of 6~MHz. A total of 25{,}530 two-dimensional DW images were collected using varying acquisition parameters, such as virtual source distances (from 10 to 20 mm) and sector angles (from 45 to 90 degrees). The imaging depth was set to 90~mm. The dataset was split into training, validation, and test sets with a ratio of 60:20:20.

All images were reconstructed using DAS beamforming with a rectangular apodization window (\( w = 1^N \)). For the following experiments, low-quality (LQ) images refer to those obtained by coherent compounding of three unfocused transmits, specifically from the leftmost, center, and rightmost steering angles. High-quality (HQ) images, used as ground truth targets during training, were reconstructed from 17 equally spaced steered DW transmits, which was found to be the best trade-off between image quality and temporal resolution.

\subsection{Implementation \& Training}

The adaptive INR and conditioner networks consist of 4 hidden layers with 128 complex-valued features per layer. For each acquisition, spatial coordinates and steering angles were normalized to the [0, 1] range and fed into the adaptive INR network. The corresponding I/Q channel data block serves as input to the local prior encoder network, producing the latent embedding used by the modulation network. The model was trained over 100 training epochs using the Adam optimizer, with an initial learning rate of 1e-4 and a batch size of 4. Our framework is implemented in PyTorch and all experiments were conducted on an NVIDIA RTX A6000 GPU.

\subsection{Method comparison}
To highlight the impact of architectural design on the expressiveness and generalization capability of INRs in ultrasound imaging, we evaluated our approach against several state-of-the-art INR-based methods originally developed for image representation tasks. The baselines include ReLU activation with positional encoding (ReLU+PE) \cite{tancik2020fourierrelupe}, Gaussian activation \cite{ramasinghe2022gaussian}, and sinusoidal activation (SIREN) \cite{sitzmann2020siren}. Since these methods do not incorporate a conditioner network that processes time-delayed channel data, their INRs are trained to infer a set of complex apodization weights, which are then element-wise multiplied with the LQ output.

We also include an alternative modulation strategy, INCODE~\cite{kazerouni2024incode}, which uses a harmonizer network to dynamically adjust parameters of the SIREN activation function. Additionally, we include two baseline feedforward networks: one that directly takes time-delayed channel data as input (FFN), similar to prior work in adaptive apodization weight prediction~\cite{luijten2020adaptive}, and another that concatenates spatial coordinates and steering angle with the channel data vector (CONCAT) \cite{li2025universal}.

\begin{figure}[H]
\centering
\includegraphics[width=0.93\textwidth]{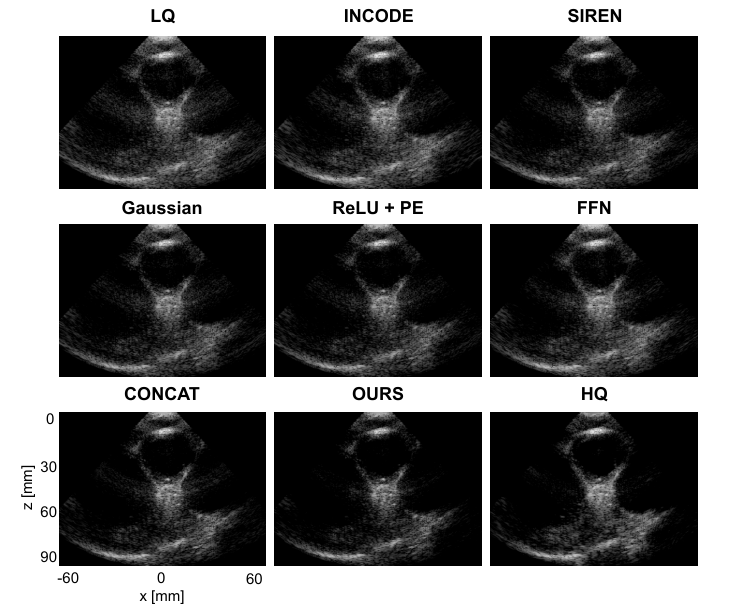}
\caption{Qualitative comparison across different methods on the \textit{in vivo} ICE dataset. The images display the aortic valve, with pronounced side lobes visible in the LQ image that are effectively suppressed by our method.} \label{fig2}
\end{figure}

\begin{figure}[H]
\centering
\includegraphics[width=0.93\textwidth]{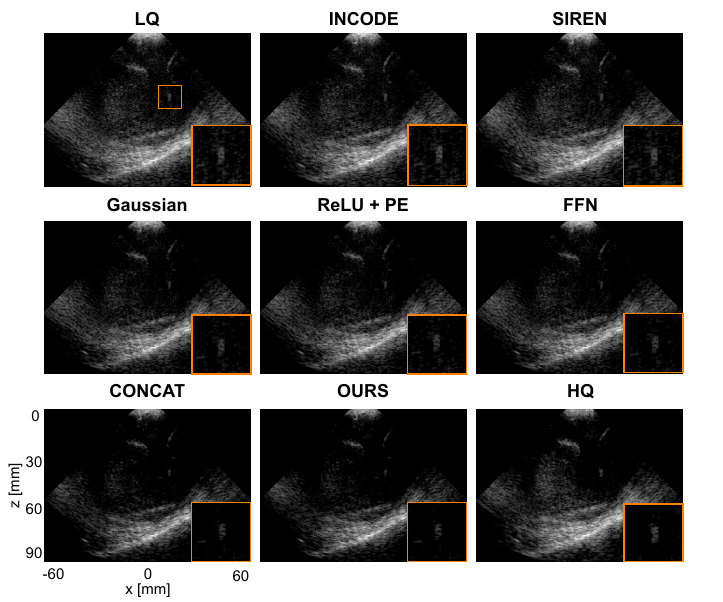}
\caption{Qualitative comparison across different methods on the \textit{in vivo} ICE dataset. The images display the tricuspid valve. The proposed framework reduces background noise and sharpens anatomical structures for improved clarity.} \label{fig3}
\end{figure}

Quantitative results are reported in Table~\ref{tab:method_comparison}. Our model achieves the best performance across peak signal-to-noise ratio (PSNR), structural similarity index (SSIM), and root-mean-square error (RMSE), reaching a PSNR of 25.34~dB and SSIM of 0.742. CONCAT emerges as the second-best method, achieving competitive scores (24.70~dB PSNR, 0.719 SSIM). Qualitative comparisons shown in Fig.~\ref{fig2} and Fig.~\ref{fig3} further confirm that our method produces sharper anatomical boundaries and substantially reduces noise artifacts compared to all baselines.

\begin{table}[ht]
\centering
\caption{Performance comparison across different methods using PSNR, SSIM, and RMSE. Best and second-best results are in \textbf{boldface} and \underline{underlined}, respectively.}
{\scriptsize
\begin{tabular}{l>{\centering}p{1.2cm}>{\centering}p{1.2cm}>{\centering}p{1.2cm}>{\centering}p{1.5cm}>{\centering}p{1.3cm}>{\centering}p{1.2cm}>{\centering}p{1.5cm}>{\centering\arraybackslash}p{1.2cm}}
\toprule
\textbf{Metric}
& \textbf{LQ}
& \textbf{FFN}
& \textbf{CONCAT}
& \makecell{\textbf{ReLU}\\\textbf{+ PE}}
& \textbf{Gaussian}
& \textbf{SIREN}
& \textbf{INCODE}
& \textbf{Ours} \\
\midrule
PSNR ↑  & 22.10 & 24.01 & \underline{24.70} & 23.82 & 23.51 & 23.48 & 23.46 & \textbf{25.34} \\
SSIM ↑  & 0.631 & 0.690 & \underline{0.719} & 0.682 & 0.670 & 0.669 & 0.668 & \textbf{0.742} \\
RMSE ↓  & 0.112 & 0.082 & \underline{0.074} & 0.078 & 0.086 & 0.087 & 0.088 & \textbf{0.069} \\
\bottomrule
\end{tabular}
}
\label{tab:method_comparison}
\end{table}

\subsection{Ablation study}
To evaluate the contribution of each architectural component in our framework, we conducted an ablation study by selectively removing or modifying individual modules. The results are summarized in Table~\ref{tab:ablation_study}. Removing the local prior encoder, i.e., directly feeding the raw I/Q channel data to the conditioner network, leads to a performance drop (PSNR: 24.20), indicating that conditioning on locally extracted features is crucial for accurate reconstruction. Excluding the conditioner network, and thus removing input-adaptive modulation, further degrades performance (PSNR: 23.82), confirming the importance of dynamic modulation in enhancing the model’s generalization capabilities.

Replacing the Gabor wavelet activation with the complex-valued Cardioid activation results in moderate performance (PSNR: 24.91), demonstrating that while Cardioid is beneficial for complex-valued signals, Gabor activation is more effective at capturing high-frequency signals necessary for high-fidelity reconstruction. Overall, the framework incorporating all components (local feature encoder, conditioner network, and Gabor wavelet activation) achieves the best performance (PSNR: 25.34).
\begin{table}[ht]
\centering
\caption{Study showing the effect of removing key components (feature encoder, conditioner network and Gabor wavelet activation) from the proposed framework.}
{\scriptsize
\begin{tabularx}{\textwidth}{l*{5}{>{\centering\arraybackslash}X}}
\toprule
\textbf{Metric} & \textbf{LQ} & \textbf{Encoder} & \textbf{Conditioner} & \textbf{Gabor} & \textbf{Ours} \\
\midrule
PSNR ↑  & 22.10 & 24.20 & 23.82 & \underline{24.91} & \textbf{25.34} \\
SSIM ↑  & 0.631 & 0.712 & 0.699 & \underline{0.729} & \textbf{0.742} \\
RMSE ↓  & 0.112 & 0.079 & 0.083 & \underline{0.073} & \textbf{0.069} \\
\bottomrule
\end{tabularx}
}
\label{tab:ablation_study}
\end{table}

\section{Conclusion}

While implicit neural representations (INRs) have shown great promise in various domains, their potential in medical imaging remains largely underexplored \cite{Molaei_2023_ICCV}. In this work, we introduced a novel modulated INR framework tailored for high-quality ultrasound image reconstruction. Our architecture employs a complex-valued INR parameterized by Gabor wavelet activations and modulated by a conditioner network. The conditioner network effectively modulates the INR network by making use of a local prior embedding extracted from the time-delayed I/Q data, enabling continuous and data-adaptive signal representation.

Comprehensive experiments on an \textit{in vivo} ICE ultrasound dataset demonstrate that our method significantly outperforms state-of-the-art INR baselines across both qualitative and quantitative evaluation. Our ablation study underscores the importance of our design choices, particularly the complex Gabor activation and adaptive modulation, in enhancing reconstruction fidelity. We believe these findings not only highlight the advantages of INR-based modeling for ultrasound, but also point toward broader opportunities for INR frameworks in medical imaging.

%
%
%
\bibliographystyle{splncs04}
\bibliography{references}

\end{document}